\documentclass{osa-article}
\usepackage{amsmath,amsfonts}      
\usepackage{braket}
\usepackage{graphicx, nicefrac}
\journal{osajournal}
\articletype{Research Article}
\usepackage{pdfpages}

\begin{document}

\title{Gigahertz-Bandwidth Optical Memory in Pr$^{3+}$:Y$_2$SiO$_5$}

\author{M. Nicolle\authormark{1, 2} J. N. Becker\authormark{1, 3}, C. Weinzetl\authormark{1}, I. A. Walmsley\authormark{1, 3}, P. M. Ledingham\authormark{1, 4}}

\address{\authormark{1}Clarendon Laboratory, University of Oxford, Parks Road, Oxford, OX1 3PU, United Kingdom\\
\authormark{2}Quantum Engineering Technology Labs, H. H. Wills Physics Laboratory \& Department of Electrical and Electronic Engineering, University of Bristol, BS8 1FD, United Kingdom\\
\authormark{3}QOLS, Blackett Laboratory, Imperial College London, London SW7 2BW, United Kingdom\\
\authormark{4}Department of Physics and Astronomy, University of Southampton, Southampton SO17 1BJ, United Kingdom}

\email{\authormark{*}Corresponding author: P.Ledingham@soton.ac.uk} 

\begin{abstract}
We experimentally study a broadband implementation of the atomic frequency comb (AFC) rephasing protocol with a cryogenically cooled Pr$^{3+}$:Y$_2$SiO$_5$ crystal. To allow for storage of broadband pulses, we explore a novel regime where the input photonic bandwidth closely matches the inhomogeneous broadening of the material $(\sim5\,\textrm{GHz})$, thereby significantly exceeding  the hyperfine ground and excited state splitting $(\sim10\,\textrm{MHz})$. Through an investigation of different AFC preparation parameters, we measure a maximum efficiency of $10\%$ after a rephasing time of $12.5\,$ns. With a suboptimal AFC, we witness up to 12 rephased temporal modes.
\end{abstract}

Quantum memories are crucial components of secure quantum communication networks \cite{Gisin2007}. Their key function is to  store and recall arbitrary quantum states of light on demand in an efficient and faithful fashion. This capability allows one to perform multiplexing, thereby enabling the synchronisation of non-deterministic events, such as the generation, distribution, and distillation of entanglement throughout the quantum network \cite{Briegel1998,Simon2007d,Kimble2008,Wehner2018}. For effective multiplexing, a storage time $\tau$ much greater than the inverse clock-rate of the system is required, therefore a large time-bandwidth product $\delta\tau$, where $\delta$ is the acceptance bandwidth of the quantum memory, is essential. For future quantum photonic networks, a high clock-rate is desirable, therefore placing a strict requirement on the acceptance bandwidth of the quantum memory.

Many quantum memory protocols have been proposed (e.g. electromagnetically induced transparency, gradient echo memory, controlled reversible inhomogeneous broadening, off-resonant Raman) and demonstrated with various material platforms (e.g. trapped atoms, warm atomic vapours, rare-earth ion doped crystals) over the last decades, see \cite{Bussieres2013a, Heshami2016} for recent reviews. One of the promising protocol-platform combinations is the atomic frequency comb (AFC) quantum memory and cryogenically-cooled praseodymium doped yttrium oxyorthosilicate - Pr$^{3+}$:Y$_2$SiO$_5$. Like all rare earth ions,  Pr$^{3+}$ is characterised by a partially filled 4f shell spatially located within the full 5s and 5p shells, resulting in relatively strong optical transitions with narrow homogeneous linewidths $(\sim\,\textrm{kHz})$ even when embedded in a solid \cite{Equall1995}. The most commonly utilised transition, the $^3H_4\rightarrow~^1D_2$ zero-phonon line with wavelength of $605.977$nm, has an absorption coefficient of $\alpha \sim 20\,\mathrm{cm}^{-1}$ for doping levels $\sim0.05\%$, an excited state lifetime of $164\mu$s, a coherence time of $111\mu$s, and a crystal-field induced inhomogeneous broadening on the order of $\sim10\,\textrm{GHz}$ \cite{Equall1995,Nilsson2004}. The ground state coherence (population) lifetime is on the order of $0.5$ms ($100$s) \cite{Ham1998}. This combination of high optical depth (OD), broadband absorption, and long coherence times make Pr$^{3+}$:Y$_2$SiO$_5$ an ideal quantum memory platform, leading to many impressive experimental demonstrations. Heralded single photons \cite{Rielander2014}, frequency-multiplexed single photons \cite{Seri2019}, and orbital-angular-momentum-encoded single photons \cite{Hua2019} have been stored using the AFC protocol. The spin-wave AFC protocol has been used to store and recall heralded single photons on demand \cite{Seri2017}. The controlled reversible inhomogenous broadening protocol has been used to store weak coherent states with $69\%$ efficiency \cite{Hedges2010}. Using external magnetic fields to operate at zero first-order Zeeman shift points \cite{Fraval2004} and dynamical decoupling techniques \cite{Fraval2005}, strong coherent states have been stored and recalled $42$s later employing electromagnetically induced transparency \cite{Heinze2013}.

Most approaches in Pr$^{3+}$:Y$_2$SiO$_5$ focus on long storage times and map light to a hyperfine ground state coherence, necessarily restricting the input pulse bandwidth to be narrow enough ($\sim\mathrm{MHz}$) to address individual transitions. So far, what is missing is a broadband implementation of a quantum memory with  Pr$^{3+}$:Y$_2$SiO$_5$. Broadband AFC demonstrations so far include Th \cite{Saglamyurek2011}, Er \cite{Jin2015}, with great potential for such a demonstration in Yb \cite{Businger2020}. The potentially long storage times of Pr$^{3+}$:Y$_2$SiO$_5$ combined with the broad acceptance bandwidth give potential for unprecedented time-bandwidth-products. To maximise the bandwidth, we here adopt a different approach compared to previous studies, utilising a significant portion of inhomogeneous broadened line to demonstrate for the first time a few-GHz bandwidth AFC optical memory in a Pr$^{3+}$:Y$_2$SiO$_5$ crystal.
The AFC scheme involves the coherent mapping of a light pulse into an ensemble of two-level atoms, where the atoms have been arranged into a series of absorbing peaks with a frequency separation of $\Delta$ and width $\gamma$. A collective coherence is established between the ground state  $|g\rangle$ and excited state $|e\rangle$ of the form $ \frac{1}{\sqrt{N}} \sum_{j=1}^Ne^{i\delta_jt}e^{-i\vec{k}_p\cdot\vec{z}_j}\ket{g_1,\dots,e_j,\dots,g_N}$ where $N$ is the total number of atoms, $\vec{k}_p$ is the input photon wave vector, $\vec{z}_j$ is the $j^\mathrm{th}$ atom position and $\delta_j$ is the detuning of the $j^\mathrm{th}$ atom with respect to resonance. This collective state rapidly dephases as each term in the sum accumulates a phase $e^{i\delta_jt}$. In the case where $\gamma \ll \Delta$, the detuning can be approximated as $\delta_j \approx m_j\Delta$, where $m_j$ are integers with the total number of $m_j$ being the number of absorbing peaks. This results in a rephasing of the collective state at a time $\tau = 2\pi/\Delta$ with a corresponding coherent photon-echo re-emission of the light \cite{Afzelius2009}. The favourable spectral-hole-burning properties of rare earth-ion-doped media, in particular Pr$^{3+}$:Y$_2$SiO$_5$, allows for the easy creation of atomic frequency comb structures. 

\begin{figure}[tbp]
\centering
\fbox{\includegraphics[width=0.7\linewidth]{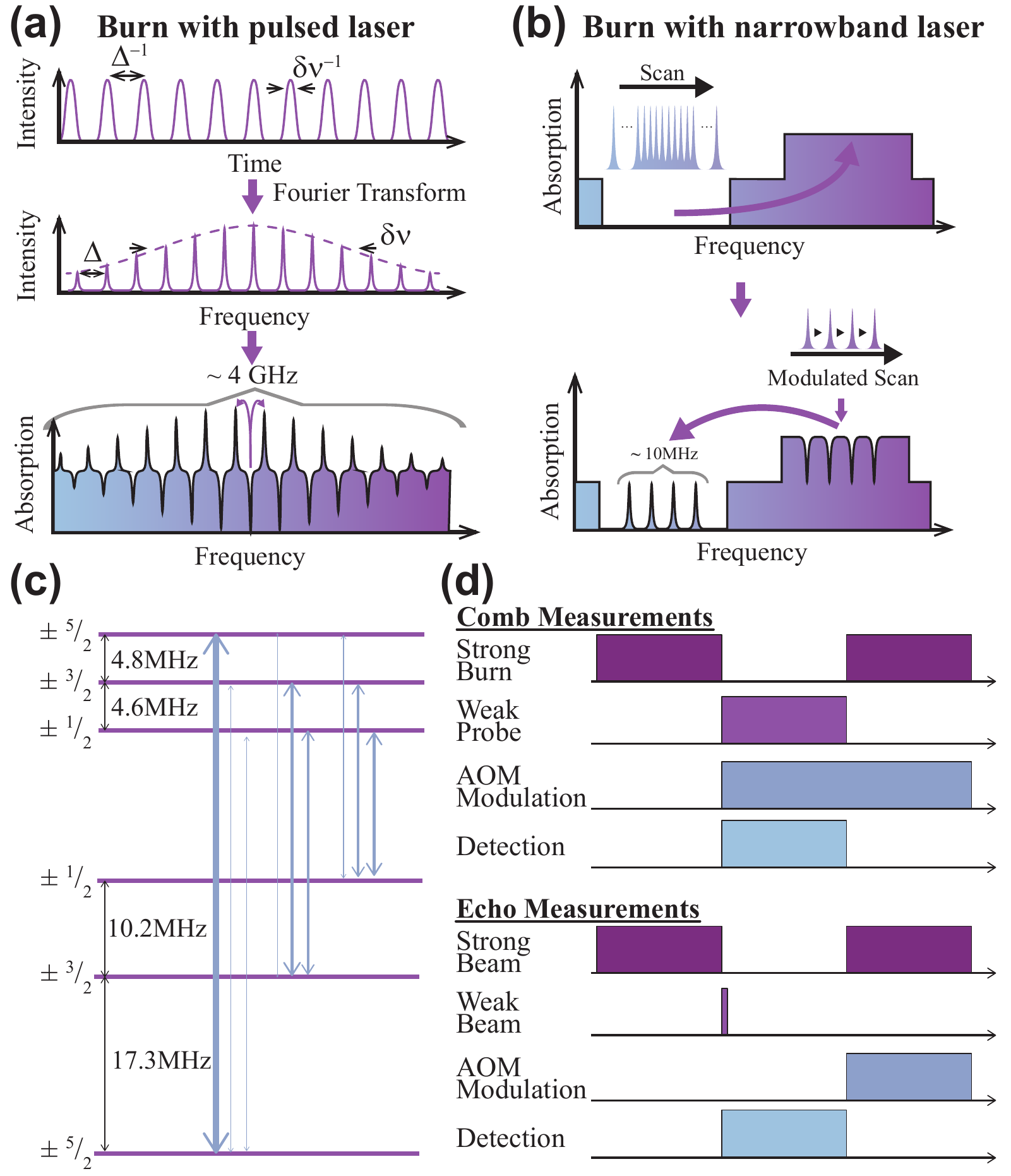}}
\caption{a) Broadband-AFC preparation scheme using an optical frequency comb. b) Typical AFC preparation using narrowband laser. c) Level scheme of the  $^3H_4(1) \rightarrow~^1D_2(1)$ transition in Pr$^{3+}$:Y$_2$SiO$_5$. d) Experimental sequences used for AFC imaging as well as echo measurements.}
\label{fig:one}
\end{figure}

In our approach, we use an optical frequency comb created by a train of 70 ps-long optical pulses from a synchronously pumped dye laser to simultaneously holeburn an AFC across a large proportion of the inhomogeneously broadened ensemble. Spectral holes are created with a spacing of $2\pi\,\times\,80\,\mathrm{MHz}$, matching the repetition rate of the laser, resulting in $\tau\,=\,12.5\,\mathrm{ns}$. Each tooth of the optical comb hereby optically pumps population via the hyperfine transitions in the fraction of atoms it addresses within the inhomogeneously broadened line. Population eventually accumulates in the $\pm\nicefrac{5}{2}$ state due to the hyperfine transition strengths (cf. Fig.\,\ref{fig:one}b), creating antiholes in between the holes and giving the comb tooth a well-defined substructure (see below). In contrast, a typical AFC preparation would utilise atoms within a narrow spectral range and starts with creating a spectral pit by sweeping a narrow-band laser across the hyperfine transitions and emptying out the $\pm\nicefrac{1}{2}$ and $\pm\nicefrac{3}{2}$ ground states. Population in the form of comb teeth is then transferred back into the pit by performing a modulated scan across the previously created antihole region \cite{Gundogan2012, Afzelius2010}.

To generate the AFC we first apply a strong pulse train to the ensemble known as the `burn'. A small fraction of the light, referred to as the `probe', can then be sent to either a scanning Fabry-Perot interferometer (FSR=1.5\,\textrm{GHz}, F=1500) to create a frequency-tunable, quasi-CW probe beam to image the comb structure or sent to a Pockels cell to pick a single signal pulse that can be read into the previously created AFC, the echoes of which can be monitored. A double-pass acousto-optic modulator (AOM) is used to rapidly frequency modulate the beam by $\pm40\,\textrm{MHz}$ during the probe period to avoid any comb structure in the probe. At the end of each experimental sequence, another strong pulse train is applied to the ensemble with the AOM modulation switched on to reset the population distribution. Switching between the individual beam paths is accomplished using motorized shutters. The sample is kept at 2.5K inside a closed-cycle helium cryostat. Light transmitted through the sample is detected using a Si avalanche photodiode (APD). An additional shutter in front of the APD protects it during the preparation reset protocols. The experimental sequences used to image the combs as well as to record photon echoes are shown in Fig.\,\ref{fig:one}c. A detailed experimental set-up is in the supplementary material.

We investigate comb characteristics as a function of burn time and power, as well as the effect of detuning from the center of the inhomogeneous line (i.e. effective OD of the ensemble). Across all measurements, we observe minimum hole widths of about 25\,\textrm{MHz}, limited by the frequency jitter of the laser, which was not actively stabilized for these experiments. Two distinct effects which lead to broadening of the comb have been observed: 1) Fluctuations in center frequency lead to a shift of the entire optical comb. 2) A change in laser repetition rate causes a breathing motion of the optical comb where teeth in the center of the comb shift less than teeth further away. Both effects can, in principle, be overcome by using a more optimized, actively stabilized laser system.

\begin{figure*}[htbp]
\centering
\fbox{\includegraphics[width=1\textwidth]{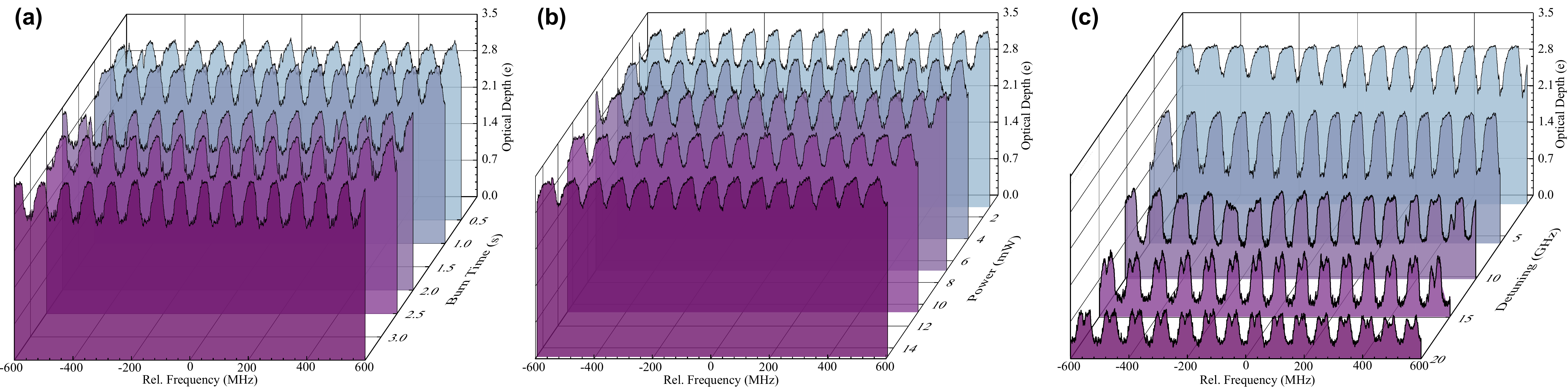}}
\caption{Broadband atomic frequency comb structures depending on a) Burn time $T_B$ i.e. length of the applied preparation pulse train ($P_B$=1.1mW). b) Burn power of the preparation pulse train $P_B$ ($T_B$=2s). c) Detuning from inhomogeneous line center ($T_B$=2s, $P_B$=1mW).}
\label{fig:two}
\end{figure*}

As shown in Fig.\,\ref{fig:two}a, after an  initial increase with burn time, the comb contrast plateaus at around 30\% after 2s ($P_B=1.1mW$). For longer burn times frequency instabilities of the laser become dominant, broadening the comb but not further increasing its contrast. Similarly, we find an optimum in burn power around 1\,\textrm{mW}. For lower powers, the pumping takes too long and the laser frequency and repetition rate instabilities become more noticeable, broadening the comb teeth. Higher powers lead to power broadening of the comb.

We investigate the comb shape as a function of detuning from the center of the inhomogeneous line. We find that, up to a detuning of about 15\,\textrm{GHz} the contrast first increases, while the peak OD decreases following the ensemble density dictated by the inhomogeneous line shape. This is attributed to the increased effective optical pumping efficiency due to the reduced OD. At larger detunings however, it is not possible to burn even deeper holes while the peak OD further decreases, reducing the overall contrast again. This indicates that a steady state of optical pumping rate and effective atomic relaxation rates is reached. We attribute this to the width of the spectral holes which are wider than the difference of ground and excited state hyperfine splittings ($\sim17\,\textrm{MHz}$). This inevitably leads to repumping between the ground hyperfine states, thus limiting the comb contrast. As validated by simulations shown in Fig.\,\ref{fig:four}a, addressing of multiple hyperfine states simultaneously also results in the asymmetric shape of the comb teeth. Simulation details can be found in the supplemental material. Both the contrast and the comb shape can be further improved by using an actively stabilized laser system, which would allow to burn holes with width narrower than 17\,\textrm{MHz} and allowing the holes to be burnt down to OD$\,=\,0$, as previously demonstrated in narrowband AFC protocols.

\begin{figure}[tbp]
\centering
\fbox{\includegraphics[width=0.65\linewidth]{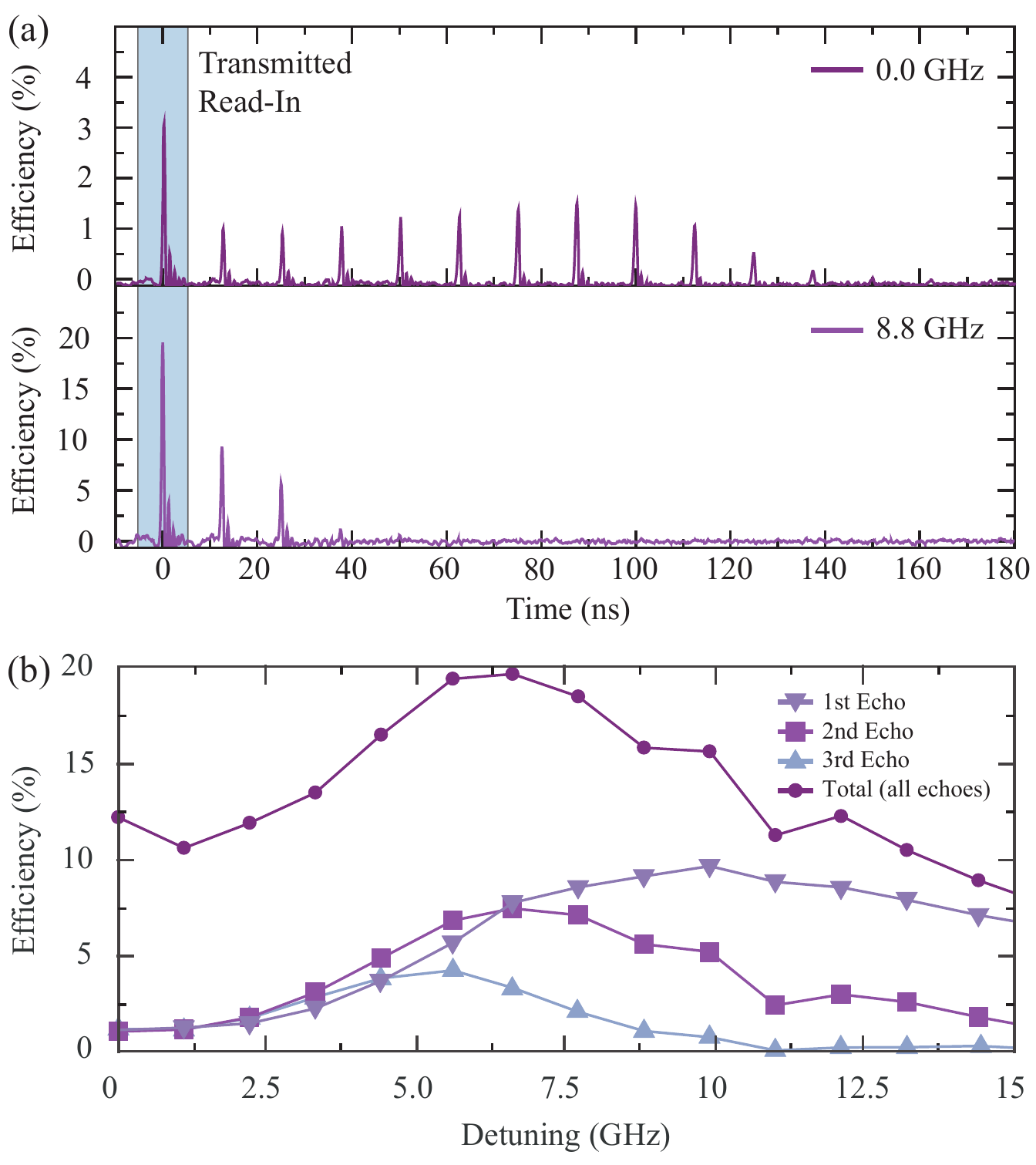}}
\caption{Photon echoes recorded with broadband AFCs. a) Time traces for 0 and 8.8\,GHz detuning from the inhomogeneous line center (burn time=2s, burn power=1mW). b) Efficiencies of the first, second and third echoes as well as total efficiency (sum of all visible echoes) for different detunings}
\label{fig:three}
\end{figure}

As a final step, we investigate the performance of the comb as a broadband AFC memory by preparing the AFC at different detunings, absorbing a strong classical pulse into this structure and observing the re-emitted photon echoes. Ideally, the AFC fully absorbs the input pulse and re-emits it entirely at the first echo time \cite{Afzelius2009}. It was possible to observe photon echoes for all combs shown in Fig.\,\ref{fig:two}c. Figure \ref{fig:three}a  shows two time traces recorded at 0 and 8.8\,GHz detuning. While the absorption efficiency at 0\,GHz is significantly higher due to the higher OD, the less ideal comb structure leads to the appearance of at least 12 visible echoes. In contrast, a better comb contrast at 8.8\,GHz allows re-emission largely within two echoes. As shown in Fig.\,\ref{fig:three}b, the relative emission into the first echo increases further with the comb contrast up to about 15\,GHz. However, the reduction in OD quickly lowers the overall (all echoes) absorption efficiency. The maximum emission efficiency into the first echo amounts to $\eta^{(1)}_{AFC}\approx 10$\% while the maximum total efficiency is $\eta^{(\mathrm{tot})}_{AFC}\approx 20$\%. Moreover, for traces in which multiple echoes are visible, we observe a complex decay envelope (in contrast to the expected mono-exponential decay) with a second local echo maximum around $100\,\mathrm{ns}$. This again is related to the addressing of multiple hyperfine states simultaneously, evidenced by the simulations shown in Fig.\,\ref{fig:four}b for which we simulated multiple combs with individual amplitudes and phases which are detuned by the hyperfine splittings. These combs can interfere with each other and create a more complex output echo trace. Simulation details can be found in the supplemental material.

\begin{figure}[tbp]
\centering
\fbox{\includegraphics[width=0.6\linewidth]{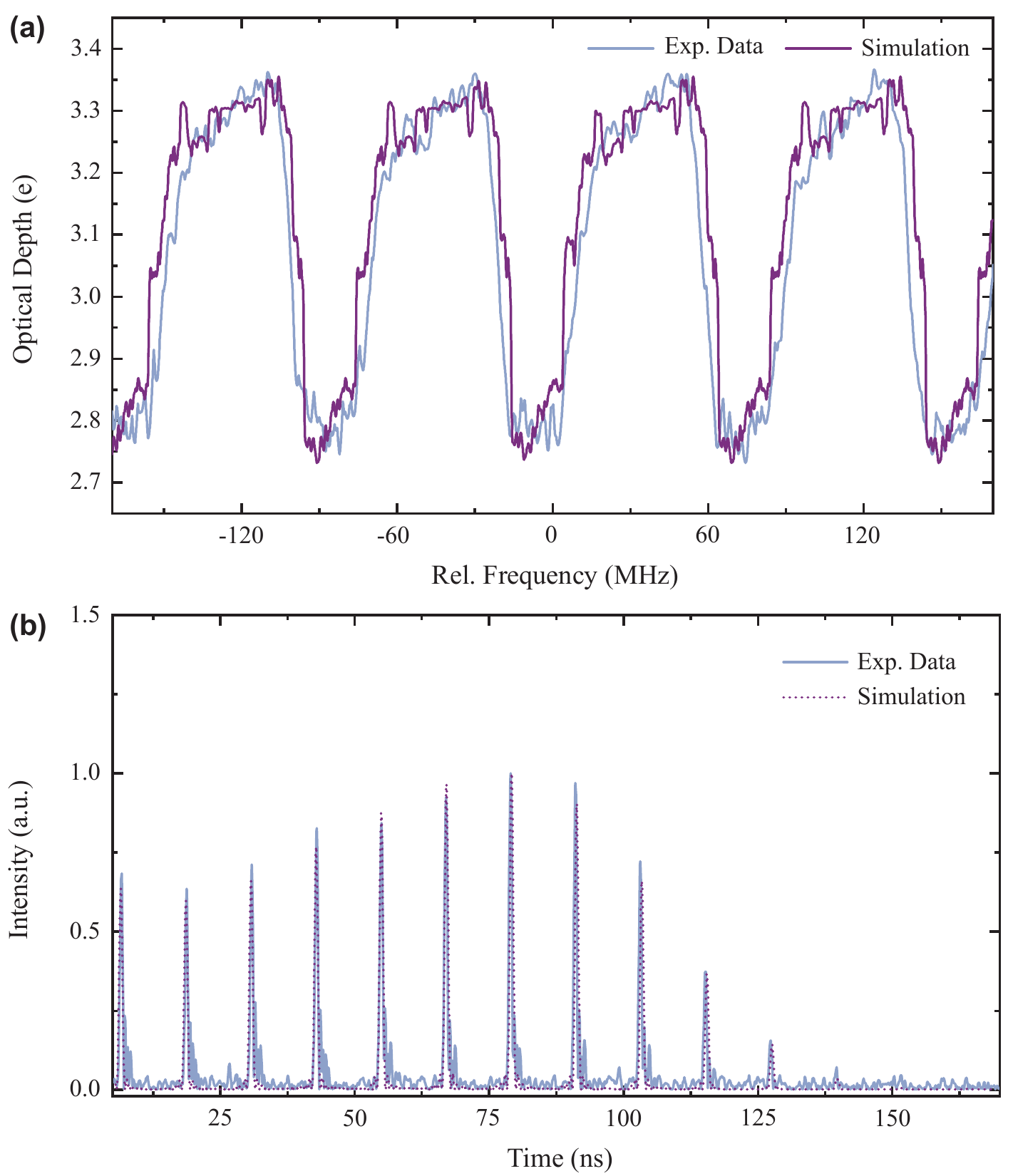}}
\caption{Simulations. a) Comparison of simulated and measured  comb shapes. b) Comparison of simulated and measured photon echoes. Comb preparation for a) and b): 0\,GHz detuning, $T_B=2s$, $P_B=1.1mW$. Details about simulations can be found in the supplemental material. }
\label{fig:four}
\end{figure}

In conclusion, we have demonstrated for the first time the preparation of broadband AFCs in Pr$^{3+}$:Y$_2$SiO$_5$ using a novel technique employing optical frequency comb generated by a broadband modelocked laser. We evaluated their performance to absorb and rephase classical broadband pulses with a bandwidth of $\sim4\,\textrm{GHz}$ in a photon echo protocol, finding maximum AFC efficiencies of $\eta^{(1)}_{AFC}\approx 10\%$, as well as the total AFC output at the $20\,\%$ level which is at a similar performance to the state-of-the-art in single pass AFC \cite{Sabooni2010, Kutluer2019}. These AFC efficiencies could be further improved by optimizing the comb preparation. For example, actively stabilizing the frequency and repetition rate of the ultrafast laser would enable longer burn times, leading to narrower comb teeth and deeper spectral holes. Alternatively, preparation using a narrowband laser in conjunction with an electro-optic modulator generating multiple higher order sidebands \cite{Seri2019} may be feasible. Additionally, increasing the (effective) sample OD, for example by using stoichiometric samples \cite{Ahlefeldt2016}, waveguide structures \cite{Seri2018, Dutta2020} and impedance-matched cavities \cite{Sabooni2013} can further enhance the absorption efficiency. 
Finally, even though a spin-wave AFC protocol to achieve long-term, on-demand storage is not compatible with our broadband AFC approach (as it requires optical access to individual long-lived ground-state hyperfine levels \cite{Afzelius2010}), achieving microsecond storage times may be feasible by combining our work with the recently demonstrated on-demand AFC read-out controlled by the Stark-effect \cite{Horvath2020}. Another approach would be to use a two-photon absorption protocol, e.g. the off-resonant cascaded absorption (ORCA) quantum memory protocol, recently demonstrated by our group \cite{Kaczmarek2018}. ORCA requires a three-level system in a ladder configuration and such a system could be formed in Pr$^{3+}$ with the $^3H_4$, $^3F_3$ and $^1D_2$ states \cite{Mazzera2011}. The scheme would work as follows: after first preparing the AFC in the ground state $^3H_4$, one could map a single photon detuned from the $^3H_4\rightarrow^3F_3$ transition into a coherence between the ground ($^3H_4$) and excited ($^1D_2$) state by a two-photon ladder process mediated with an off-resonant control pulse driving the $^3F_3\rightarrow^1D_2$ transition. The application of a second control pulse at precisely the AFC rephasing time will read out the memory. If the control pulse is not applied or is mistimed, the memory will not be phase-matched for read-out. In fact, the coherence can now be read out at any time multiple of the rephasing time thereby transforming the AFC protocol to an on-demand quantum memory protocol modulo $\tau$ \cite{Rubio2018}. A further advantage with this approach is that the $^3H_4\rightarrow^3F_3$ transition is around $1550\,\mathrm{nm}$ which would enable Pr$^{3+}$:Y$_2$SiO$_5$ to directly interface to telecommunication C-band quantum networks.

\begin{backmatter}
\bmsection{Funding} UK Engineering and Physical Sciences Research Council (EPSRC) Standard Grant No. EP/J000051/1, Programme Grant No. EP/K034480/1, the EPSRC Hub for Networked Quantum Information Technologies (NQIT), ERC Advanced Grant (MOQUACINO).

\bmsection{Acknowledgments} We thank Margherita Mazzera for useful discussions and reading of the manuscript, and Hugues de Riedmatten for supplying the Pr$^{3+}$:Y$_2$SiO$_5$ crystal.

\bmsection{Disclosures} The authors declare no conflicts of interest.

\end{backmatter}
\bibliography{TheBib}

\newpage
\begin{center}
\Large{
\textbf{Supplement: Gigahertz-Bandwidth Optical Memory in Pr$^{3+}$:Y$_2$SiO$_5$}}\\
\normalsize
M. Nicolle$^{1,2}$, J. N. Becker$^{1,3}$, C. Weinzetl$^{1}$, I. A. Walmsley$^{1,3}$, P. M. Ledingham $^{1,*}$\\
$^{1}$Clarendon Laboratory, University of Oxford, Parks Road, Oxford, OX1 3PU, United Kingdom\\
$^{2}$Quantum Engineering Technology Labs, H. H. Wills Physics Laboratory \& Department of Electrical and Electronic Engineering, University of Bristol, BS8 1FD, United Kingdom\\
$^{3}$QOLS, Blackett Laboratory, Imperial College London, London SW7 2BW, United Kingdom\\
$^{4}$Department of Physics and Astronomy, University of Southampton, Southampton SO17 1BJ, United Kingdom\\
$^{*}$Corresponding author: P.Ledingham@soton.ac.uk\\
\end{center}
\section{Experimental Setup}
For atomic frequency comb (AFC) preparation we use an optical frequency comb created by a train of 70ps-long optical pulses at 606\,nm from a synchronously pumped dye laser (Sirah Gropius) using Rhodamine 6G in ethylene glycol. The laser is pumped by a frequency-doubled Yb-doped fiber laser (NKT aeroPulse PS with APE Emerald Engine) with a fundamental pulse width of 5\,ps and repetition rate of 80\,MHz. This pulse train is sent through a 80\,MHz double-pass AOM setup with variable-frequency driver to optionally modulate the laser frequency by $\pm40$MHz to wash out the comb structure. This is used both to reset the AFC at the end of each cycle as well as to create a quasi-CW probe beam in conjunction with a scanning Fabry-Perot cavity (Thorlabs, FSR=1.5\,GHz, F=1500) to image the AFC structure. Alternatively, the beam can be sent through a Pockels cell (BME Bergmann, 1\,kHz) to pick a single pulse which can be used as a signal pulse to be read into the AFC. The sample is a commercially available Pr$^{3+}$:Y$_2$SiO$_5$ crystal (0.05\% Pr$^{3+}$, length=3\,mm) kept at 2.5\,K in a closed-cycle helium cryostat (Oxford Instruments Optistat Dry) using custom-made copper heat shields. Light transmitted through the sample is detected by an AC-coupled 1\,GHZ Si avalanche photodiode (Menlo Systems APD210). Switching between the individual beam paths is accomplished using mechanical shutters controlled by a digital delay generator (Stanford Research Systems DG645) via an Arduino interface and sychronized to the trigger signal generated by the pulsed laser.

\begin{figure}[h]
\centering
\includegraphics[width=\linewidth]{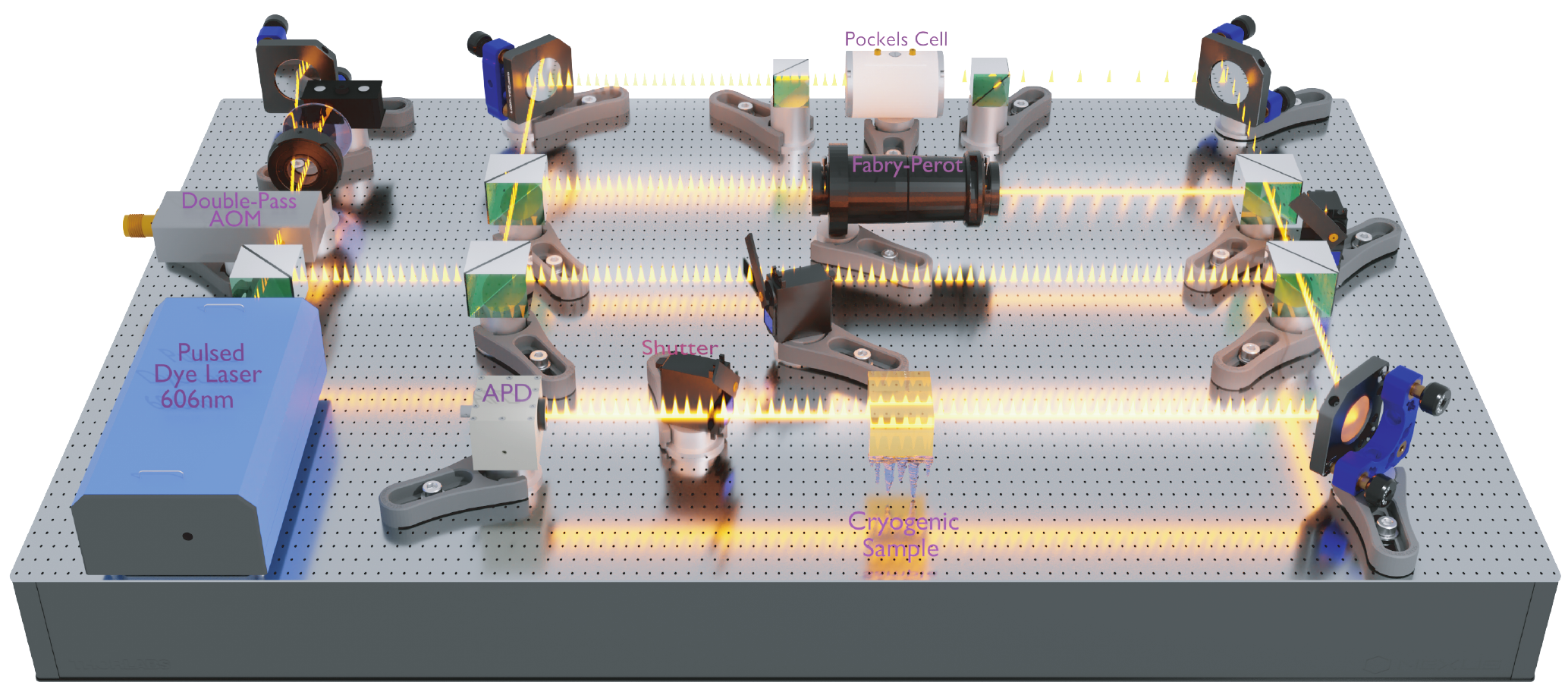}
\caption{Experimental setup used to conduct the experiments described in the main text.}
\label{fig:Sone}
\end{figure}
\section{Simulations}

\subsection{Comb shape}

We model the hole burning process in order to calculate the simulated comb profile shown in Figure 4a of the main text. As the inhomogeneous linewidth is orders of magnitude wider than the linewidth of each individual ion, a monochromatic probe can address up to nine different transitions in different ions shifted in frequency with respect to each other. Optically pumping the ensemble with a single frequency then modifies the optical transmission at a multitude of other frequencies by redistributing the relative hyperfine ground state populations. We calculate this resulting absorption profile in our simulations using the relative oscillator strengths for the optical transitions between the $^3\mathrm{H}_4$ and $^1\mathrm{D}_2$ hyperfine manifolds (\cite{Nilsson2004}.

The burning laser is modelled as a collection of narrowband probes separated by 80 MHz which jitters by a maximum of 40 MHz around a central frequency, inspired from measurements of the laser spectrum. Each of the burning frequencies is taken into account in order to compute the redistribution of population along the inhomogeneous line and the resulting atomic frequency comb profile. Within this framework, relaxation of the hyperfine states is neglected.

\subsection{Echo shape}

In the AFC protocol, the temporal profile of the rephasing echoes is determined by the Fourier transform of the atomic spectral distribution. For example, if the AFC profile is composed of a set of evenly spaced delta-functions in frequency, the output will be a pulse train in the time domain. What is observed in Figure 4b of the main text is not a simple pulse train, but one where the amplitude of each pulse is forms an interesting profile. The profile is down to the finite peak width of the comb teeth leading to an exponential decay of the coherence, as well as the interference between the multiple combs corresponding to different hyperfine levels of Pr$^{3+}$:Y$_2$SiO$_5$. In our simulation we assume that by the end of the comb preparation process, almost all of the ions are pumped into the $\pm5/2$ hyperfine ground state, which is coupled to three optical transitions. These transitions are modelled as three distinct atomic frequency combs with equal spacing but detuned with respect to each other by an amount equal to the hyperfine state splitting. The overall envelope of the echo pulse train is then simulated by calculating the beat signal between the frequencies composing the full atomic spectral distribution, and is found to accurately match the observed data. This envelope further decays exponentially according to a time constant defined by the inverse of the AFC peak width.
\end{document}